# Solvable RSOS models based on the dilute BWM algebra

Uwe Grimm *
Instituut voor Theoretische Fysica
Universiteit van Amsterdam
Valckenierstraat 65
1018 XE Amsterdam
The Netherlands

and

S. Ole Warnaar †
Mathematics Department
University of Melbourne
Parkville
Victoria 3052
Australia

**Abstract**

In this paper we present representations of the recently introduced dilute Birman–Wenzl–Murakami algebra. These representations, labelled by the level-$l$ $\mathrm{B}_n^{(1)}$, $\mathrm{C}_n^{(1)}$ and $\mathrm{D}_n^{(1)}$ affine Lie algebras, are Baxterized to yield solutions to the Yang–Baxter equation. The thus obtained critical solvable models are RSOS counterparts of the, respectively, $\mathrm{D}_{n+1}^{(2)}$, $A_{2n}^{(2)}$ and $\mathrm{B}_n^{(1)}$ $R$-matrices of Bazhanov and Jimbo. For the $\mathrm{D}_{n+1}^{(2)}$ and $\mathrm{B}_n^{(1)}$ algebras the RSOS models are new. An elliptic extension which solves the Yang–Baxter equation is given for all three series of dilute RSOS models.

*e-mail: `grimm@phys.uva.nl`
†e-mail: `warnaar@mundoe.maths.mu.oz.au`

# 1 Introduction

During the past decade, there has been spectacular progress in the understanding of the behaviour of two-dimensional statistical systems at or close to their critical point. Indisputably, a lot has been learned from conformal field theory which, for instance, led to a deeper understanding of the concept of universality. Another main reason for progress is the enormous amount of examples which can be treated by analytic methods, notably the lattice models. There, a most prominent role is played by the Yang–Baxter equation (YBE) being a sufficient condition for the solvability of a model in the sense of commuting transfer matrices [1]. Furthermore, the study of solutions to the YBE also had considerable impact in other fields of mathematics and physics, the most important probably being the introduction of quantum groups, but also the quite unexpected implications for the theory of knot and link invariants is worth mentioning.

Although a lot is known about solutions of the YBE, no complete classification has been established so far. The basic solutions (in the sense that others can be built from these by the fusion procedure) fall into two mutually dual classes, corresponding to vertex models on one hand and so-called solid–on–solid (SOS) models and their restricted counterparts (RSOS models) on the other. They are at least partially characterized by the classification scheme of affine Lie algebras, and the corresponding vertex models, for all but the exceptional algebras, have been obtained by Bazhanov and by Jimbo [2]. Related RSOS models were constructed by Andrews, Baxter and Forrester [3] for $A_1^{(1)}$, and by Jimbo, Miwa and Okado [4] for the non-twisted affine Lie algebras $A_n^{(1)}$, $B_n^{(1)}$, $C_n^{(1)}$ and $D_n^{(1)}$. RSOS models corresponding to the twisted algebras $A_n^{(2)}$ have been considered by Kuniba [5].

However, in general there can be several distinct series of solvable RSOS models related to the same affine Lie algebra. In particular, a second series of so-called dilute models related to $A_2^{(2)}$ was constructed [6, 7] which recently has been generalized to the general rank case $A_n^{(2)}$ in [8]. These models are quite different from Kuniba's models, and in particular they include models which are solvable at the critical temperature in the presence of a symmetry-breaking field.

The purpose of this paper is the construction of two new infinite series of solvable models related to $B_n^{(1)}$ and $D_{n+1}^{(2)}$. In our working we also include the $A_{2n}^{(2)}$ models of [8] as they can be treated completely analogously. The construction is based on an algebraic approach, considering a "dilution" of the representations of the Birman–Wenzl–Murakami (BWM) algebra [9] which underly the critical $B_n^{(1)}$, $C_n^{(1)}$ and $D_n^{(1)}$ RSOS models of ref. [4].

The paper is organized as follows. In the subsequent section, we commence with a complete description of the dilute Birman–Wenzl–Murakami (dBWM) algebra. The defining relations are motivated by making extensive use of the diagrammatic interpretation of the dilute algebra which generalizes the ordinary BWM algebra by allowing for "vacancies". In sec. 3, we construct three infinite series of representations based on representations of the BWM algebra related to the critical $B_n^{(1)}$, $C_n^{(1)}$ and $D_n^{(1)}$ RSOS models of Jimbo, Miwa and Okado [4], and hence are labelled by the level-$l$ dominant integral weights of the corresponding affine Lie algebras. Following ref. [10], these representations are then Baxterized in sec. 4, yielding three series of critical solvable models being related to respectively the $D^{(2)}_{n+1}$, $A^{(2)}_{2n}$ and $B^{(1)}_{n}$ $R$ matrices of Bazhanov and Jimbo [2]. The models can be extended off criticality while preserving solvability, and in sec. 5 the corresponding elliptic face weights are presented explicitly. The $B^{(1)}_{n}$ and $D^{(2)}_{n+1}$ models obtained in this way are new whereas the $A^{(2)}_{2n}$ models have recently been found by similar methods in [8]. In sec. 6, we summarize and discuss our results, and point out some possible generalizations of the work described in this paper. Finally, two identities used in the main text are proved in the appendix.

# 2 The dilute BWM algebra

In the context of two-dimensional solvable models, the so-called braid–monoid algebras [9, 11, 12] have frequently been considered in the literature. On the one hand, representations of these algebras might give rise to new solutions of the Yang–Baxter equation, this approach to the construction of solvable models being known as *Baxterization* [13] (see also [14]). On the other hand, they allow for a graphical interpretation in terms of operators acting on arrays of *strands* or *strings* (see for instance [12]) in such a way that the relations in the algebra are essentially equivalent to continuous deformations of the corresponding diagrams. Hence, they play an important role in providing a connection between solvable lattice models and the theory of knot and link invariants [11, 12, 15, 16, 17].

The notion of a dilute braid–monoid algebra naturally emerged from the investigation of the dilute A–D–E models [6, 7, 18, 19], which can be described in terms of a dilute generalization of the Temperley–Lieb (TL) algebra [20]. A formal and more general definition has been given in ref. [21] where it is regarded as a special case of a two–colour generalization of the braid–monoid algebra. Although this is natural from the point of view of generalizing to multi–colour algebras, the dilute braid–monoid algebra has several simplifying features when considered as a one–colour algebra with vacancies. Since in this paper, as well as in a number of recent publications [10, 22, 8], the dilute algebras (and in particular the dilute BWM algebra) play a prominent role, we will in this section define this algebra in its true dilute form as opposed to its earlier non-dilute two–colour definition in refs. [21, 10, 22].

However, before we commence to do so, we want to stress that the definition given below is of course nothing else than a formal way to state the relations which are implied in the diagrammatic approach used in refs. [8, 23]. This is based on the idea that one considers diagrams similar to those of the usual braid–monoid algebras (see e.g., [12]), but in addition one allows for vacancies, i.e., for the absence of strings. In other words, we now consider two possible states at each position: either there is a string, which is represented by a line, or there is no string. In contrast to refs. [7, 18, 23, 8] where the vacancies were represented by dotted or dashed lines, we will now leave them out completely.

All relations we now impose on our dilute algebra are those obtained from the usual braid–monoid relations [12] by discarding an arbitrary number of strings. Of course, not all the relations obtained in this way are independent, and consequently the choice of an independent subset of relations is by no means unique. The definition given below is based on one particular choice of defining relations which, to our taste, yields the simplest expressions.

## 2.1 Graphical interpretation of the dilute algebra

In view of the above remarks, we will follow a somewhat twisted approach and introduce the dilute algebras by actually discussing the graphical interpretation of their generators and relations, prior to a formal definition. This has the advantage that the defining relations, which would otherwise seem to come out of the blue, acquire a clear interpretation in terms of the diagrams presented below, which cannot be visualized in the equations themselves. In fact, the best way to understand the relations is to look at the corresponding pictures, but for the sake of space we have to restrict ourselves to a few examples.

Clearly, the algebra is characterized by an integer $N+1$ which in the case of a dilute algebra gives the number of possible positions of strings (as opposed to the actual number of strings in the non-dilute case). In our graphical interpretation, an element of the algebra is therefore represented by two rows of $N+1$ points (symbolized by small circles) each and their connections by strings. If two strings cross, we distinguish between two possibilities ("over" and "under" crossing) as in the non-dilute case, and one can think of the diagrams as projections from curves which are embedded in three-dimensional space. Multiplication of algebra elements corresponds to concatenation of the corresponding diagrams, where we use the convention that for $A \cdot B$ the diagram of $A$ is placed on top of that of $B$. Vertical dotted lines between the points represent the identity of the algebra, indicating that this includes either of the allowed states (string or vacancy) at the corresponding positions. The identity $\mathcal{I}$ can thus be represented graphically as

$\mathcal{I}$ =

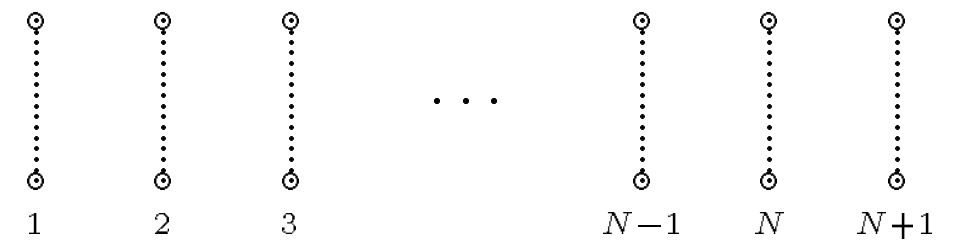

. (2.1)

The generators $s_j$ and $v_j$ $(1 \leq j \leq N+1)$ "creating" a string, respectively a vacancy, at position $j$ are depicted

$s_j$ =

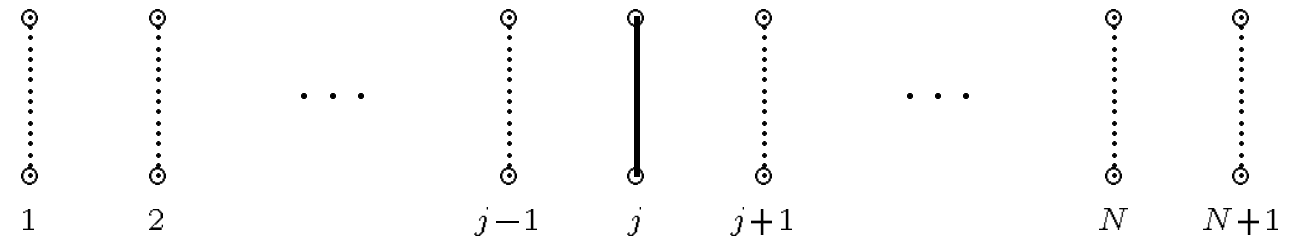

(2.2)

$v_j$ =

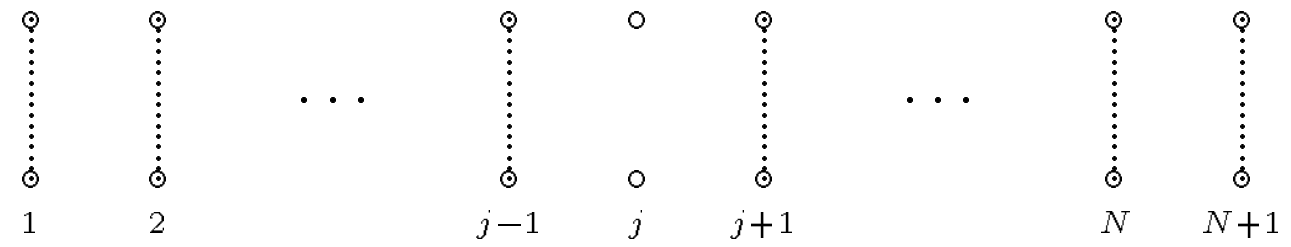

. (2.3)

These two operators are to be understood as orthogonal projectors, their product being zero. Hence any diagram with sinks or sources of strings corresponds to zero in the algebra.

From now on, we will in general concentrate on the non-trivial part of the diagrams, neglecting any position where the corresponding elements of the dilute algebra act as the identity. The remaining generators $b_j^{\pm}$, $e_j$, $(/)_j$, $(\backslash)_j$, $({}^{\cup})_j$ and $({}_{\cap})_j$ $(1 \leq j \leq N)$ act non-trivially at positions $j$ and $j+1$ only, with the following "reduced" diagrams:

$b_j^+ = (\times)_j =$ 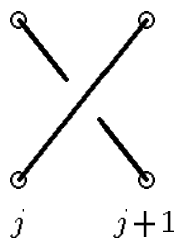

$b_j^- = (\times)_j =$ 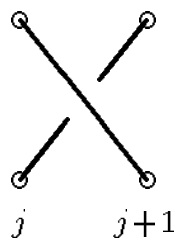

$e_j = (\asymp)_j =$ 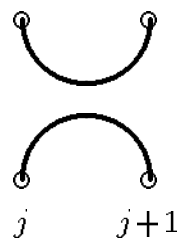

$(/)_j =$ 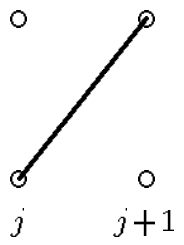

$(\backslash)_j =$ 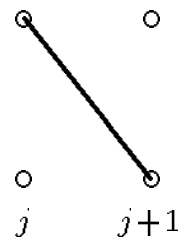

(2.4)

$({}^{\cup})_j =$ 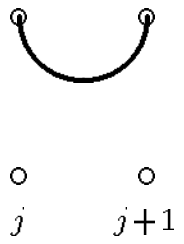

$({}_{\cap})_j =$ 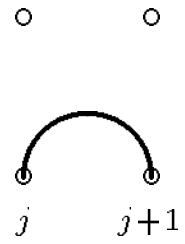

.

With the above notation, we denote $\mathcal{B}_j = \{\, b_j^{\pm}\,,\, e_j \,\}$ and $\mathcal{D}_j = \{(/)_j, (\backslash)_j, ({}^{\cup})_j, ({}_{\cap})_j\}$.

Relations in the algebra are now related to continuous deformations of the corresponding diagrams. As in the usual non-dilute case [12], closed loops yield a factor $\sqrt{Q}$ and removing a "twist" (corresponding to a Reidemeister move I) yields a factor $\omega$ or $\omega^{-1}$. For instance

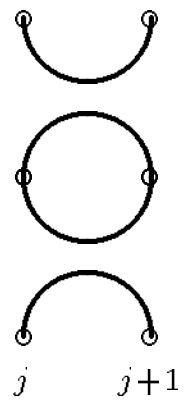

$= \sqrt{Q}$ 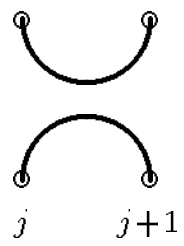

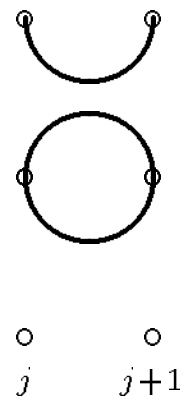

$= \sqrt{Q}$ 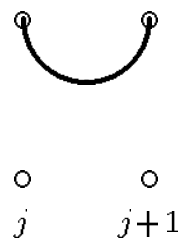

(2.5)

and

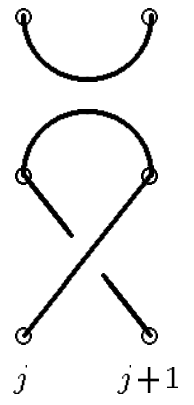

$= \omega$ 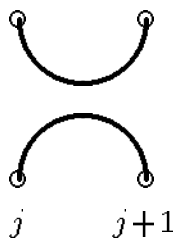

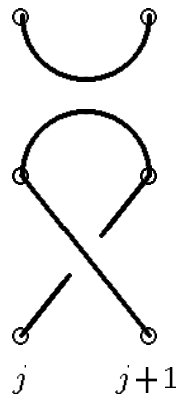

$= \omega^{-1}$ 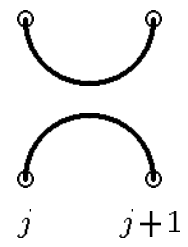

. (2.6)

Algebra elements whose diagrams are related by *regular isotopy* [24] (i.e., by Reidemeister moves II and III alone) are identical.

Similarly, one finds relations involving other generators. Two typical examples are

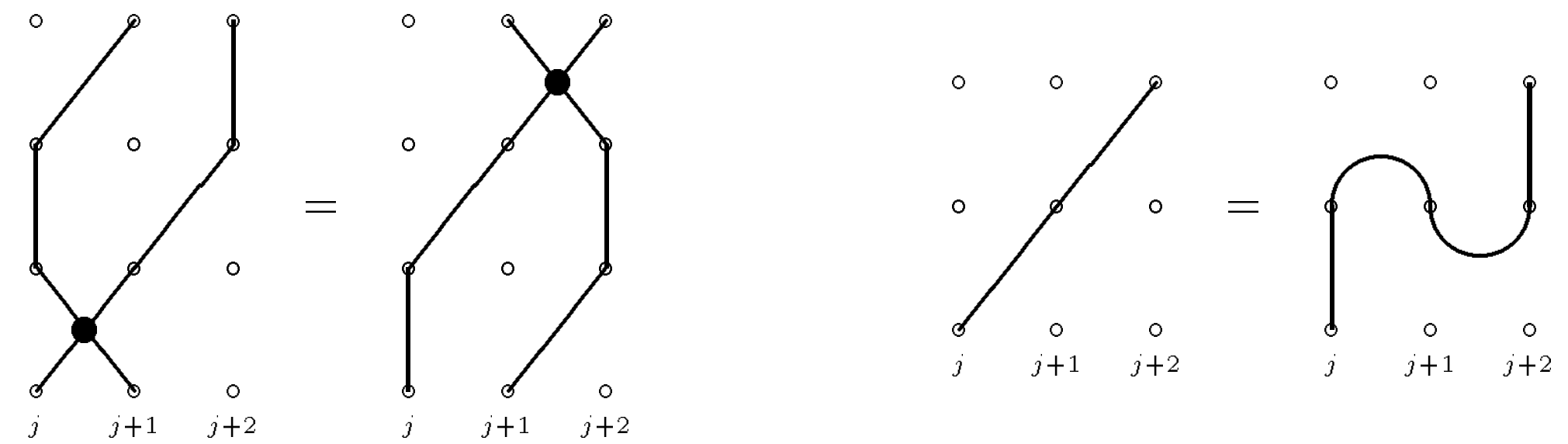

(2.7)

Here the two vertices with solid circle $\bullet$ represent an arbitrary product of elements in $\mathcal{B}_j$ and $\mathcal{B}_{j+1}$, respectively.

We note that, instead of the standard notation $b_j$ and $b_j^{-1}$ for the braids generators $(\times)_j$ and $(\times)_j$, we use the slightly different notation $b_j^+$ and $b_j^-$. This is to reflect the fact that in our case $(\times)_j$ and $(\times)_j$ are *not* truly inverse to each other, their product $b_j^+ \, b_j^- = b_j^- \, b_j^+ = s_j \, s_{j+1}$ being a projector (or, if one prefers, the identity in the corresponding subspace):

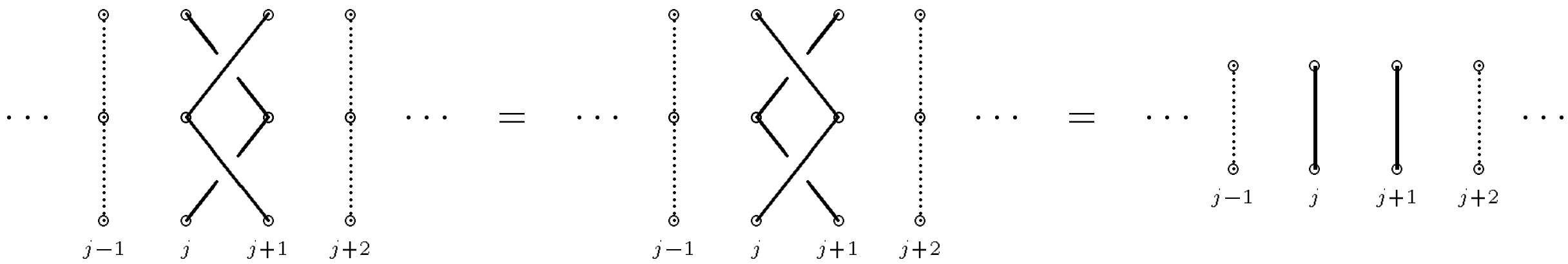

## 2.2 Definition of dBWM algebra

Let us now start to give a complete definition of what we mean by a *dilute Birman–Wenzl–Murakami* (dBWM) *algebra* or, more generally, by a *dilute braid–monoid algebra* (compare ref. [21]). It is an algebra with identity $\mathcal{I}$ generated by $s_j$ and $v_j$ (with $1 \leq j \leq N+1$), together with the generators in $\bigcup_{j=1}^{N}(\mathcal{B}_j \cup \mathcal{D}_j)$ and two central elements ("constants") $\sqrt{Q}$ and $\omega$, subject to a list of relations given below.

To start with, we assume that the action of the generators is local in the sense that they commute whenever their indices sufficiently differ, i.e.,

$$\begin{aligned} \mathcal{P}_j \, \tilde{\mathcal{P}}_k &= \tilde{\mathcal{P}}_k \, \mathcal{P}_j && \text{for } j \neq k \\ \mathcal{P}_j \, \mathcal{O}_k &= \mathcal{O}_k \, \mathcal{P}_j && \text{for } j \neq k, k+1 \\ \mathcal{O}_j \, \tilde{\mathcal{O}}_k &= \tilde{\mathcal{O}}_k \, \mathcal{O}_j && \text{for } |j-k| > 1, \end{aligned} \tag{2.8}$$

where the generators $\mathcal{P}_j$ and $\tilde{\mathcal{P}}_j$ are in the set $\{ s_j , v_j \}$ and $\mathcal{O}_j , \tilde{\mathcal{O}}_j \in \mathcal{B}_j \cup \mathcal{D}_j$.

The generators $s_j$ and $v_j$ fulfill the relations

$$s_j + v_j = \mathcal{I} \qquad {s_j}^2 = s_j \ , \tag{2.9}$$

which imply ${v_j}^2 = v_j$ and $s_j\, v_j = v_j\, s_j = 0$, and hence are orthogonal projectors. For the sake of brevity and for later convenience, we also introduce projectors acting on two neighbouring sites,

$$
\begin{aligned}
I_j \;=\; (\rangle\langle)_j \;&=\; s_j\, s_{j+1} \qquad & (\rangle\;\;)_j \;&=\; s_j\, v_{j+1} \\
(\;\;\langle)_j \;&=\; v_j\, s_{j+1} \qquad & (\quad)_j \;&=\; v_j\, v_{j+1}\;.
\end{aligned}
\tag{2.10}
$$

Here, we use slight bends in our symbols to make it easier to distinguish the two "mixed" projectors.

The compatibility relations between the projectors and the remaining generators can now conveniently be summarized in the following form:

$$
\begin{aligned}
(\rangle\langle)_j\, \mathcal{O}_j\, (\rangle\langle)_j \;&=\; \mathcal{O}_j \qquad && \text{for } \mathcal{O}_j \in \mathcal{B}_j \\
(\;\;\langle)_j (\diagup)_j (\rangle\;\;)_j \;&=\; (\diagup)_j \qquad && (\rangle\;\;)_j (\diagdown)_j (\;\;\langle)_j \;=\; (\diagdown)_j \\
(\rangle\langle)_j (^{\smile})_j (\quad)_j \;&=\; (^{\smile})_j \qquad && (\quad)_j (_{\frown})_j (\rangle\langle)_j \;=\; (_{\frown})_j\;.
\end{aligned}
\tag{2.11}
$$

One can think of these relations as defining the "external legs" (i.e., the positions of incoming and outgoing strings) of our generators.

To justify the name *dilute braid–monoid algebra*, we require that $b_j^+$, $b_j^-$, and $e_j$ $(1 \leq j \leq N)$, generate a braid–monoid subalgebra[1], and hence that the following relations hold [12]:

$$
\begin{aligned}
b_j^+\, b_j^- \;&=\; b_j^-\, b_j^+ \;=\; I_j \\
b_j^+\, b_{j+1}^+\, b_j^+ \;&=\; b_{j+1}^+\, b_j^+\, b_{j+1}^+ \\
{e_j}^2 \;&=\; \sqrt{Q}\; e_j \\
e_j\, e_{j\pm1}\, e_j \;&=\; e_j\, I_{j\pm1} \\
b_j^+\, e_j \;&=\; e_j\, b_j^+ \;=\; \omega\; e_j \\
b_j^+\, b_{j\pm1}^+\, e_j \;&=\; e_{j\pm1}\, b_j^+\, b_{j\pm1}^+ \;=\; e_{j\pm1}\, e_j
\end{aligned}
\tag{2.12}
$$

or, in our graphical notation,

$$
\begin{aligned}
(\diagdown\!\!\!\!\diagup)_j (\diagup\!\!\!\!\diagdown)_j \;&=\; (\diagup\!\!\!\!\diagdown)_j (\diagdown\!\!\!\!\diagup)_j \;=\; (\rangle\langle)_j \\
(\diagdown\!\!\!\!\diagup)_j (\diagdown\!\!\!\!\diagup)_{j+1} (\diagdown\!\!\!\!\diagup)_j \;&=\; (\diagdown\!\!\!\!\diagup)_{j+1} (\diagdown\!\!\!\!\diagup)_j (\diagdown\!\!\!\!\diagup)_{j+1} \\
{(^{\smile}_{\frown})_j}^2 \;&=\; \sqrt{Q}\, (^{\smile}_{\frown})_j \\
(^{\smile}_{\frown})_j (^{\smile}_{\frown})_{j\pm1} (^{\smile}_{\frown})_j \;&=\; (^{\smile}_{\frown})_j (\rangle\langle)_{j\pm1}
\end{aligned}
\tag{2.13}
$$

[1]To be precise we should say that the operators $b_j^+\, I$, $b_j^-\, I$ and $e_j\, I$, with $I = \prod_j I_j$, generate a braid–monoid algebra on the completely occupied subspace.

$$(\times)_j(\asymp)_j \;=\; (\asymp)_j(\times)_j \;=\; \omega\,(\asymp)_j$$

$$(\times)_j(\times)_{j\pm1}(\asymp)_j \;=\; (\asymp)_{j\pm1}(\times)_j(\times)_{j\pm1} \;=\; (\asymp)_{j\pm1}(\asymp)_j \;.$$

Finally, we demand the following set of relations for the mixed generators to be satisfied

$$\begin{array}{rclrcl}
(/)_j(\backslash)_j &=& (\;\langle)_j & (\backslash)_j(/)_j &=& (\rangle\;)_j \\
(\frown)_j(\asymp)_j &=& \sqrt{Q}\,(\frown)_j & (\asymp)_j(\smile)_j &=& \sqrt{Q}\,(\smile)_j \\
(\frown)_j(\smile)_j &=& \sqrt{Q}\,(\;\;)_j & (\smile)_j(\frown)_j &=& (\asymp)_j \\
(/)_j(/)_{j+1}(\smile)_j &=& (\;\langle)_j(\smile)_{j+1} & (\frown)_j(\backslash)_{j+1}(\backslash)_j &=& (\frown)_{j+1}(\;\langle)_j \\
(/)_{j+1}(/)_j &=& (\frown)_j(\smile)_{j+1} & (\backslash)_j(\backslash)_{j+1} &=& (\frown)_{j+1}(\smile)_j \\
(/)_j(/)_{j+1}(\times)_j &=& (\times)_{j+1}(/)_j(/)_{j+1} \;. & & &
\end{array} \tag{2.14}$$

If all the relations listed above are fulfilled, we name the corresponding algebra a *dilute braid–monoid algebra*. This definition is consistent with the "two–colour" definition of ref. [21] since, assuming $Q \neq 0$, all relations of ref. [21] follow from those listed above. In order to obtain what we call a *dilute* BWM (dBWM) *algebra*, we require in addition that the braid–monoid subalgebra on the completely occupied subspace (see previous footnote) is in fact a BWM algebra, i.e., the following polynomial reduction relations hold [9]:

$$\left(b_j^+ - q^{-1} I_j\right)\left(b_j^+ + q\, I_j\right)\left(b_j^+ - \omega\, I_j\right) \;=\; 0 \tag{2.15}$$

$$e_j \;=\; \frac{\omega^{-1}}{q-q^{-1}}\left(b_j^+ - q^{-1} I_j\right)\left(b_j^+ + q\, I_j\right) \;=\; I_j \,+\, \frac{b_j^+ \,-\, b_j^-}{q-q^{-1}} \tag{2.16}$$

$$\sqrt{Q} \;=\; 1 \,+\, \frac{\omega-\omega^{-1}}{q-q^{-1}} \;. \tag{2.17}$$

Before we proceed to discuss several representations of the dBWM algebra, let us make some more remarks. Define $B_j$, $B_j^{-1}$ and $E_j$ $(1 \leq j \leq N)$ by [21]

$$\begin{aligned}
B_j &= (\times)_j - (/)_j - (\backslash)_j + \sigma\,(\;\;)_j \\
B_j^{-1} &= (\bar{\times})_j - (/)_j - (\backslash)_j + \sigma\,(\;\;)_j \\
E_j &= (\asymp)_j + (\smile)_j + (\frown)_j + (\;\;)_j
\end{aligned} \tag{2.18}$$

where $\sigma^2 = 1$. It then follows from the defining relations of the dilute algebra that $B_j$ and $B_j^{-1}$ satisfy the braid group relations (and in particular are really inverse to each other) and the polynomial equation

$$(B_j - \mathcal{I}\,)\,(B_j + \mathcal{I}\,)\,(B_j - q^{-1}\,\mathcal{I}\,)\,(B_j + q\,\mathcal{I}\,)\,(B_j - \omega\,\mathcal{I}\,) \;=\; 0 \;. \tag{2.19}$$

Moreover, the $E_j$ generate a TL algebra with

$${E_j}^2 \;=\; \sqrt{\tilde{Q}}\, E_j \;=\; (\sqrt{Q}+1)\, E_j \;, \tag{2.20}$$

and most of the braid–monoid relations are fulfilled, see ref. [21] for details. However, the algebra generated by $B_j$, $B_j^{-1}$ and $E_j$ is *not* a braid–monoid algebra in the usual sense, as

$$\begin{aligned} B_j E_j &= \omega \left(\substack{\cup \\ \cap}\right)_j + \omega \left({}^{\cup}\right)_j + \sigma \left({}_{\cap}\right)_j + \sigma \left(\;\;\right)_j \\ E_j B_j &= \omega \left(\substack{\cup \\ \cap}\right)_j + \sigma \left({}^{\cup}\right)_j + \omega \left({}_{\cap}\right)_j + \sigma \left(\;\;\right)_j \end{aligned} \tag{2.21}$$

and hence $B_j$ and $E_j$ do *not* commute for $\omega \neq \sigma$.

In spite of this, one can of course, via eqs. (2.18) and (2.20), define a TL model for any representation of the dBWM algebra. This way of constructing TL models is similar in spirit to that of ref. [25], where TL models were obtained from the $\mathrm{X}_n^{(1)}$ BWM representations underlying the $\mathrm{X}_n^{(1)}$ vertex and RSOS models of refs. [2, 4]. However, from eq. (2.20) it is obvious that for the restricted models one has $\tilde{Q} > 4$ in general. Hence it follows from the TL equivalence with the self-dual $\tilde{Q}$–state Potts model [20] that the models constructed in this way are not critical.

# 3 Representations of the dBWM algebra

In the following we provide three infinite families of representations of the dBWM algebra, labelled by the $\mathrm{B}_n^{(1)}$, $\mathrm{C}_n^{(1)}$ and $\mathrm{D}_n^{(1)} \equiv \mathrm{X}_n^{(1)}$ affine Lie algebras at level $l$. These representations are obtained by diluting the ordinary BWM representations underlying the $\mathrm{X}_n^{(1)}$ RSOS models of Jimbo, Miwa and Okado [4, 26].

Some of the characteristics of such a level-$l$ $\mathrm{X}_n^{(1)}$ representation can be conveniently encoded by a graph, and before we actually present the dBWM representations, we discuss some general notions as adjacency graphs, path spaces and admissibility rules.

## 3.1 Adjacency graphs, admissibility and path spaces

Consider a set of nodes each labelled by a *height*, all nodes having distinct label. We let the heights be coordinates in some $m$ dimensional linear space. Now draw a set of *bonds* between the nodes, such that we get a connected graph $\mathcal{G}$. Between each pair of nodes we allow for at most one bond. If nodes $a$ and $b$ are connected by a bond[2] they are called *adjacent* or *admissible* and denoted by $a \sim b$. If $a \sim b$ then also $b \sim a$ (i.e., the bonds are undirected). Since $\mathcal{G}$ is connected there is always a sequence of adjacent heights $a \sim b \sim c \sim \ldots \sim f$ connecting two arbitrary nodes $a$ and $f$ of $\mathcal{G}$.

We now define the set of vectors obtained by taking all differences of adjacent nodes as $\mathcal{A}$, that is, if $a \sim b \Rightarrow \epsilon = a - b \in \mathcal{A}$. The converse of this is not necessarily true, i.e., $a - b \in \mathcal{A} \not\Rightarrow a \sim b$. Clearly, since $a \sim b \Leftrightarrow b \sim a$, each vector $\epsilon_i \in \mathcal{A}$ has an inverse $-\epsilon_i \equiv \epsilon_{-i}$ also in $\mathcal{A}$.

[2]We do not distinguish between a node and its height.

In the following we consider two types of graphs. Graphs without nodes connected to themselves and graphs with some or all nodes connected to themselves via a singe bond, referred to as a *tadpole*. For the first type of graphs we have $\mathcal{A} = \{\pm\epsilon_1, \ldots, \pm\epsilon_n\}$ and for the second $\mathcal{A} = \{0, \pm\epsilon_1, \ldots, \pm\epsilon_n\}$. For reasons which will become clear later, we rewrite this second set as $\mathcal{A} = \{\epsilon_0, \pm\epsilon_1, \ldots, \pm\epsilon_n\}$.

Now consider a random walker making an excursion on $\mathcal{G}$. Each discrete time interval the walker can either take a step along a bond or he can remain standing. The collection of all paths on $\mathcal{G}$ generated by our walker in $N+1$ time steps, where we take all nodes as possible starting points, is denoted by $\mathcal{H}_N$ and named the *path space* of $\mathcal{G}$. Elements $\{a\}$ of $\mathcal{H}_N$ are called admissible paths. In all dBWM representations to be defined in section 3.3 below, the generators $\mathcal{O}_j$ of the algebra will act in some path space $\mathcal{H}_N$ defined by an *adjacency* graph $\mathcal{G}$.

For graphs with tadpoles it, in general, does not suffice to give the sequence of heights visited by our walker, to denote a typical path $\{a\}$ in $\mathcal{H}_N$. For example, letting $b$ be a node with tadpole, the sequence $\ldots, a, b, b, c, \ldots$ could either mean that the walker took a step along the tadpole, or that he took a rest while standing on $b$. We therefore denote elements of $\mathcal{H}_N$ by explicitly writing each step taken by the walker and using the vector $\epsilon_0$ for steps along a tadpole. In this notation two typical paths $\{a\}$ and $\{a'\}$, which could not be distinguished by giving the sequences of heights only, are

$$\begin{aligned} \{a\} &= a,\, a+\epsilon_\mu,\, a+\epsilon_\mu+\epsilon_\nu,\, a+\epsilon_\mu+\epsilon_\nu,\, a+\epsilon_\mu+\epsilon_\nu+\epsilon_\tau,\, \ldots \\ \{a'\} &= a,\, a+\epsilon_\mu,\, a+\epsilon_\mu+\epsilon_\nu,\, a+\epsilon_\mu+\epsilon_\nu+\epsilon_0,\, a+\epsilon_\mu+\epsilon_\nu+\epsilon_0+\epsilon_\tau+\epsilon_0,\, \ldots \end{aligned} \tag{3.1}$$

So, in the first path the walker takes a rest after two steps and in the second path he takes a step along the tadpole. In spite of the above remarks, we on some occasions nevertheless write $\{a\} = a_0, a_1, \ldots, a_{N+1}$ to denote an arbitrary path of $\mathcal{H}_N$.

We could of course, as an alternative to the above notation, replace each node with a tadpole by a node with two tadpoles, and each node without a tadpole by a node with a single tadpole. We then would have the rule that the walker has to make a step each time interval, and we could simply add one additional vector, say $\epsilon_{\bar{0}}$ to $\mathcal{A}$. However, we feel that the approach chosen here is more natural, as the sets $\mathcal{A}$ as will be defined below have an immediate interpretation in terms of the affine Lie algebras $\mathrm{X}_n^{(1)}$. This is similar to our earlier convention for the dilute A–D–E models [7], where we defined the adjacency graphs as the Dynkin diagrams of the simply laced Lie algebras and not as dressed-up Dynkin diagrams with a tadpole at each node.

As we will see later, also in connection with the dBWM algebra the above approach is preferable, as the two possible "string"-states at each position are naturally related to our random walker. If the walker takes a step at time interval $i$ this is associated with a string at position $i$, and if the walker takes a rest at time interval $i$, this is associated with a vacancy at position $i$. Indeed, in presenting the actual representations of the dBWM algebra and in presenting the resulting solvable models, the sets $\mathcal{A}$ as chosen here admit the simplest listing of results.

## 3.2 Level-$l$ $\mathrm{X}_n^{(1)}$ adjacency graphs

With the notions introduced above, we are now in the position to define the graphs which will encode our dBWM representations. To do so, we first define the set of heights that will be the nodes of our adjacency graphs. We then give the sets $\mathcal{A}$ and finally we formulate an admissibility rule stating for which pairs of heights $a$ and $b$ with $a-b \in \mathcal{A}$ we have that $a \sim b$ and hence that $a$ and $b$ are connected by a bond.

### 3.2.1 Sets of heights

In the three series of dBWM algebra representations to be considered here, the heights $a$ are given by the level-$l$ dominant integral weights of the non-twisted affine Lie algebras $\mathrm{X}_n^{(1)} = \mathrm{B}_n^{(1)}$, $\mathrm{C}_n^{(1)}$ and $\mathrm{D}_n^{(1)}$ [27].

Letting $g$ be the dual Coxeter number, $t$ the (long root)$^2/2$ and $l$ the level of $\mathrm{X}_n^{(1)}$, we set

$$L = t(l+g), \tag{3.2}$$

with $g$ and $t$ listed in Table 1. Denoting the fundamental weights of $\mathrm{X}_n^{(1)}$ by $\Lambda_0, \ldots, \Lambda_n$, the allowed heights $a$ for the three respective series read

$$\begin{aligned}
&\mathrm{B}_n^{(1)} \begin{cases} a = (L - a_1 - a_2 - 1)\Lambda_0 + \sum_{i=1}^{n-1} (a_i - a_{i+1} - 1)\,\Lambda_i + (2a_n - 1)\Lambda_n \\[2ex] L > a_1 + a_2,\; a_1 > a_2 > \ldots > a_n > 0, \quad \text{all } a_i \in \mathbb{Z}, \text{ or all } a_i \in \mathbb{Z} + \frac{1}{2} \end{cases} \\
&\mathrm{C}_n^{(1)} \begin{cases} a = (L/2 - a_1 - 1)\Lambda_0 + \sum_{i=1}^{n-1} (a_i - a_{i+1} - 1)\,\Lambda_i + (a_n - 1)\Lambda_n \\[2ex] L/2 > a_1 > a_2 > \ldots > a_n > 0, \quad a_i \in \mathbb{Z} \end{cases} \\
&\mathrm{D}_n^{(1)} \begin{cases} a = (L - a_1 - a_2 - 1)\Lambda_0 + \sum_{i=1}^{n-1} (a_i - a_{i+1} - 1)\,\Lambda_i + (a_{n-1} + a_n - 1)\Lambda_n \\[2ex] L > a_1 + a_2,\; a_1 > a_2 > \ldots > a_n,\; a_{n-1} + a_n > 0, \quad \text{all } a_i \in \mathbb{Z}, \text{ or all } a_i \in \mathbb{Z} + \frac{1}{2}. \end{cases}
\end{aligned} \tag{3.3}$$

### 3.2.2 Admissibility rules

We now define $\mathcal{A}$ as the set of weights in the vector representation of the classical Lie algebra $\mathrm{X}_n$, and express the elements of $\mathcal{A}$ in terms of the orthonormal vectors $\epsilon_i$, $1 \le i \le n$,

| $X_n^{(1)}$ | $g$ | $t$ |
|---|---|---|
| $B_n^{(1)}$ | $2n-1$ | 1 |
| $C_n^{(1)}$ | $n+1$ | 2 |
| $D_n^{(1)}$ | $2n-2$ | 1 |

Table 1: The dual Coxeter number $g$ and the (long root)$^2/2$ of the algebra $X_n^{(1)}$.

$\langle \epsilon_i, \epsilon_j \rangle = \delta_{i,j}$,

$$\mathcal{A} = \begin{cases} \{0, \pm\epsilon_1, \dots, \pm\epsilon_n, \epsilon_0\} \equiv \{\epsilon_0, \pm\epsilon_1, \dots, \pm\epsilon_n\} & \mathrm{B}_n \\ \{\pm\epsilon_1, \dots, \pm\epsilon_n\} & \mathrm{C}_n,\ \mathrm{D}_n. \end{cases} \tag{3.4}$$

We can then write the classical part of the weights, denoted by $\bar{\Lambda}_i$, as

$$\begin{aligned} &\mathrm{B}_n^{(1)} \begin{cases} \bar{\Lambda}_i = \epsilon_1 + \ldots + \epsilon_i & 1 \le i \le n-1 \\ \quad\ = \frac{1}{2}(\epsilon_1 + \ldots + \epsilon_n) & i = n \end{cases} \\ &\mathrm{C}_n^{(1)} \quad \bar{\Lambda}_i = \epsilon_1 + \ldots + \epsilon_i \qquad 1 \le i \le n \\ &\mathrm{D}_n^{(1)} \begin{cases} \bar{\Lambda}_i = \epsilon_1 + \ldots + \epsilon_i & 1 \le i \le n-2 \\ \quad\ = \frac{1}{2}(\epsilon_1 + \ldots + \epsilon_{n-2} + \epsilon_{n-1} - \epsilon_n) & i = n-1 \\ \quad\ = \frac{1}{2}(\epsilon_1 + \ldots + \epsilon_{n-2} + \epsilon_{n-1} + \epsilon_n) & i = n. \end{cases} \end{aligned} \tag{3.5}$$

If we also introduce the symbol $\rho = \Lambda_0 + \ldots + \Lambda_n$, and set $\epsilon_{-i} = -\epsilon_i$, we get from (3.3) for the classical part of $a+\rho$: $\bar{a} + \bar{\rho} = \sum_{i=1}^n a_i \epsilon_i$, and hence $a_\mu =< a+\rho, \epsilon_\mu >$, $-n \le \mu \le n$, $\mu \ne 0$. For $\mu = 0$ we set $a_0 = -\frac{1}{2}$.

With the above definitions, we can now formulate the rules yielding the adjacency graphs $\mathcal{G}$.

> Let $V(\bar{a})$ being an irreducible $\mathrm{X}_n$ module with highest weight $\bar{a}$.
>
> The nodes $a$ and $b$ of $\mathcal{G}$ are connected by a bond iff for any Dynkin
>
> diagram automorphism $\mu$, the tensor module $V(\mu(\bar{a})) \otimes V(\bar{\Lambda}_1)$ (3.6)
>
> includes $V(\mu(\bar{b}))$.

For $\mathrm{C}_n^{(1)}$ and $\mathrm{D}_n^{(1)}$ this simply means that $a$ and $b$ are connected iff $a - b \in \mathcal{A}$. For $\mathrm{B}_n^{(1)}$ we have, in addition to this, to implement the rule that tadpoles on $a$ are absent if $a_n = \frac{1}{2}$. Some simple level-$l$ $\mathrm{X}_n^{(1)}$ adjacency graphs are shown in figure 1.

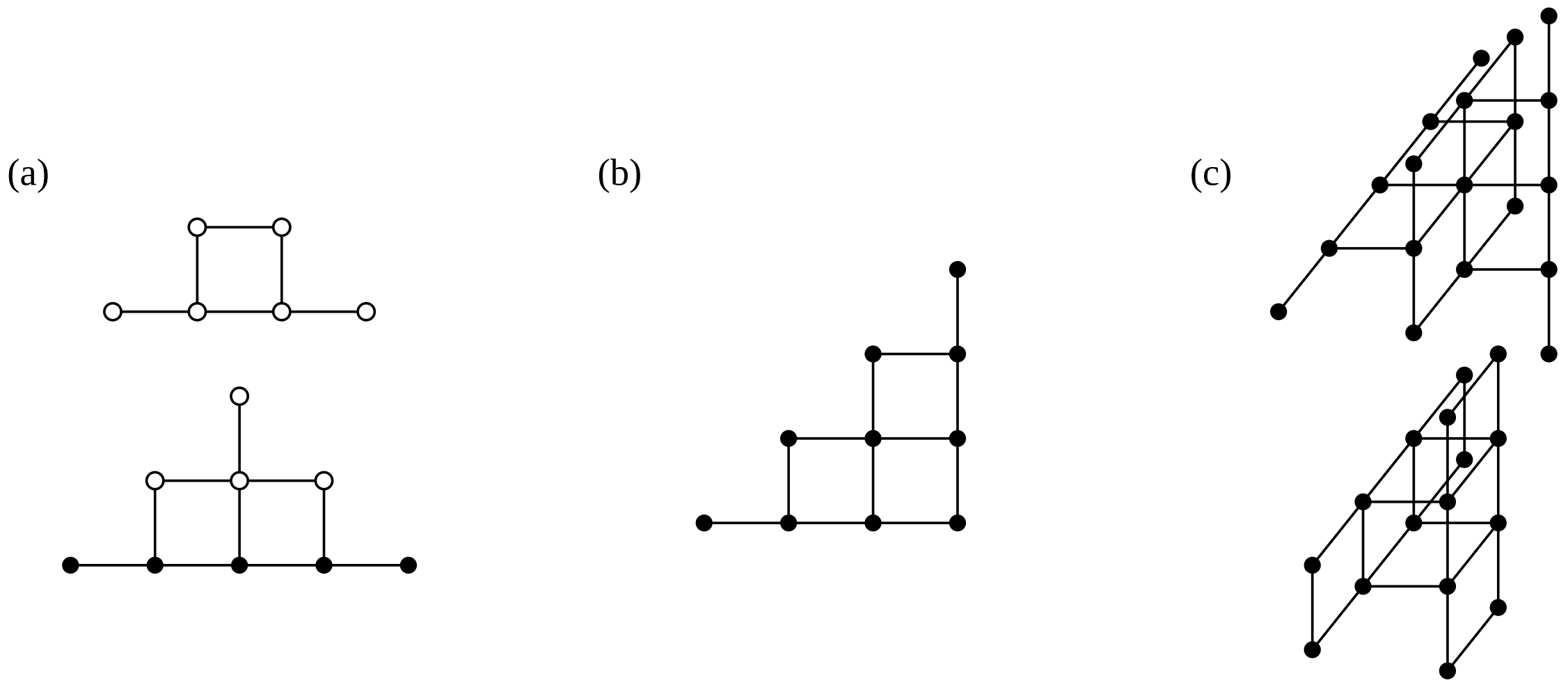

Figure 1: (a) Adjacency graphs for $B_2^{(1)}$ level 4. The top(bottom) graph has $a_i \in \mathbb{Z}(\mathbb{Z}+\frac{1}{2})$, with $a_i$ as in (3.3). The open circles denote nodes with a tadpole, i.e., $a \sim a$ and the solid circles denote nodes not connected to themselves, i.e., $a \not\sim a$. (b) Adjacency graph for $C_2^{(1)}$ level 3. (c) Adjacency graphs for $D_3^{(1)}$ level 4. The top(bottom) graph has $a_i \in \mathbb{Z}(\mathbb{Z}+\frac{1}{2})$.

### 3.3 $X_n^{(1)}$ representations of the dBWM algebra

We now list the representations of the dBWM algebra as follow from the level-$l$ $X_n^{(1)}$ adjacency graphs $\mathcal{G}$. Hereto we define the function

$$[u] = \sin\left(\frac{s\pi u}{L}\right) \qquad s \in \mathbb{Z}, \tag{3.7}$$

with $s$ and $L$ coprime, and introduce the following notation for listing the matrix elements of the dBWM generators $\mathcal{O}_j$, acting in $\mathcal{H}_N$:

$$(\mathcal{O}_j)_{\{a\},\{b\}} = \mathcal{O}\begin{pmatrix} a_{j-1} & b_j \\ a_j & a_{j+1} \end{pmatrix} \prod_{k\neq j} \delta_{a_k,b_k}. \tag{3.8}$$

We then have $X_n^{(1)}$ dBWM representations with constants

$$\begin{aligned} \omega &= \sigma\, q^{2\lambda} \\ \sqrt{Q} &= \frac{[1-2\sigma\lambda][\frac{1}{2}+\sigma\lambda]}{[\frac{1}{2}-\sigma\lambda][1]} = 1+\sigma\,\frac{[2\lambda]}{[1]} \\ q &= \mathrm{e}^{-\mathrm{i}\,\pi s/L} \end{aligned} \tag{3.9}$$

and non-zero matrix elements:

$$e\begin{pmatrix} a & a+\epsilon_\nu \\ a+\epsilon_\mu & a \end{pmatrix} = (G_{a,\mu}G_{a,\nu})^{1/2}$$

$$
\begin{aligned}
&b^{\pm}\begin{pmatrix} a & a+\epsilon_\mu \\ a+\epsilon_\mu & a+2\epsilon_\mu \end{pmatrix} = q^{\mp 1} \qquad \mu \neq 0 \\
&b^{\pm}\begin{pmatrix} a & a+\epsilon_\mu \\ a+\epsilon_\mu & a+\epsilon_\mu+\epsilon_\nu \end{pmatrix} = -q^{\pm(a_\mu - a_\nu)}\,\frac{[1]}{[a_\mu - a_\nu]} \qquad \mu \neq \pm\nu \\
&b^{\pm}\begin{pmatrix} a & a+\epsilon_\nu \\ a+\epsilon_\mu & a+\epsilon_\mu+\epsilon_\nu \end{pmatrix} = -\left(\frac{[a_\mu - a_\nu + 1][a_\mu - a_\nu - 1]}{[a_\mu - a_\nu]^2}\right)^{1/2} \qquad \mu \neq \pm\nu \\
&b^{\pm}\begin{pmatrix} a & a+\epsilon_\nu \\ a+\epsilon_\mu & a \end{pmatrix} = (G_{a,\mu}G_{a,\nu})^{1/2}\, q^{\pm(a_\mu + a_\nu + 1)}\,\frac{[1]}{[a_\mu + a_\nu + 1]} \qquad \mu \neq \nu \\
&b^{\pm}\begin{pmatrix} a & a+\epsilon_\mu \\ a+\epsilon_\mu & a \end{pmatrix} = (G_{a,\mu} - 1)\, q^{\pm(2a_\mu + 1)}\,\frac{[1]}{[2a_\mu + 1]} \qquad \mu \neq 0 \\
&\qquad\qquad = (1 - H_{a,\mu})\, q^{\pm(2a_\mu + 1)}\,\frac{[1]}{[2a_\mu - 2\lambda + 1]} \qquad\qquad (3.10) \\
&(\rangle\langle)\begin{pmatrix} a & a+\epsilon_\mu \\ a+\epsilon_\mu & a+\epsilon_\mu+\epsilon_\nu \end{pmatrix} = 1 \\
&(\rangle\ )\begin{pmatrix} a+\epsilon_\mu & a \\ a & a \end{pmatrix} = (\ \langle)\begin{pmatrix} a & a \\ a & a+\epsilon_\mu \end{pmatrix} = 1 \\
&(\ \ )\begin{pmatrix} a & a \\ a & a \end{pmatrix} = 1 \\
&(/)\begin{pmatrix} a & a \\ a+\epsilon_\mu & a+\epsilon_\mu \end{pmatrix} = (\backslash)\begin{pmatrix} a & a+\epsilon_\mu \\ a & a+\epsilon_\mu \end{pmatrix} = 1 \\
&(\frown)\begin{pmatrix} a & a \\ a+\epsilon_\mu & a \end{pmatrix} = (\smile)\begin{pmatrix} a & a+\epsilon_\mu \\ a & a \end{pmatrix} = (G_{a,\mu})^{1/2}.
\end{aligned}
$$

The functions $G_{a,\mu}$ and $H_{a,\mu}$ herein read

$$
\begin{aligned}
G_{a,\mu} &= \sigma\,\frac{h(a_\mu + 1)}{h(a_\mu)} \prod_{\nu \neq 0, \pm\mu} \frac{[a_\mu - a_\nu + 1]}{[a_\mu - a_\nu]} \qquad \mu \neq 0 \\
G_{a,0} &= 1 \qquad\qquad (3.11) \\
H_{a,\mu} &= \sum_{\nu \neq \mu} G_{a,\nu} \frac{[a_\mu + a_\nu - 2\lambda + 1]}{[a_\mu + a_\nu + 1]} + 1.
\end{aligned}
$$

The sign factor $\sigma$, the variable $\lambda$ and the function $h$ are as given in Table 2. We remark here that $\lambda$ will play the role of *crossing parameter* in the subsequently defined solvable models. The range of $\mu$ and $\nu$ in (3.10) and (3.11) is, as follows from (3.4), $\mu, \nu = 0, \pm 1, \ldots, \pm n$ for $\mathrm{B}^{(1)}_{n+1}$, and $\mu, \nu = \pm 1, \ldots, \pm n$ for $\mathrm{C}^{(1)}_n$ and $\mathrm{D}^{(1)}_n$. For the $\mathrm{B}^{(1)}_n$ case we recall that $a_0 = -\frac{1}{2}$.

The above representations based on $\mathrm{C}_n^{(1)}$ have been obtained previously in [8]. Representations dual[3] to those in (3.10) have been found in [22].

| $\mathrm{X}_n^{(1)}$ | $\lambda$ | $\sigma$ | $h(a)$ | $\mathrm{X}_m^{(p)}$ |
|---|---|---|---|---|
| $\mathrm{B}_n^{(1)}$ | $n$ | $1$ | $[a]$ | $\mathrm{D}_{n+1}^{(2)}$ |
| $\mathrm{C}_n^{(1)}$ | $n+\frac{1}{2}$ | $-1$ | $[2a]$ | $\mathrm{A}_{2n}^{(2)}$ |
| $\mathrm{D}_n^{(1)}$ | $n-\frac{1}{2}$ | $1$ | $1$ | $\mathrm{B}_n^{(1)}$ |

Table 2: Crossing parameter, sign factor and the function $h$ for the $\mathrm{X}_n^{(1)}$ representations of the dBWM algebra. In the last column we list the affine Lie algebra associated with the solvable model obtained after Baxterization. Note that $\lambda = (tg+\sigma)/2$, with $t$ and $g$ given in Table 1.

In proving the expressions (3.9)–(3.11) to be valid, it is simplest to first consider the unrestricted case where we take $L$ in (3.3) to be infinite and $L$ in (3.7) to be an arbitrary complex parameter $\neq 0$. The adjacency graphs are then infinite graphs with nodes $a$ and $b$ connected if $a-b \in \mathcal{A}$. Once the dBWM representations are shown to be correct in this infinite case, we only have to assert that (3.10) still holds in truncating $\mathcal{G}$ to its finite form, now letting $L$ be given by (3.2). This is simply a matter of case checking similar to that of [4] in their restriction of the $\mathrm{X}_n^{(1)}$ SOS models.

Proving the unrestricted case, we substitute (3.10), into the defining relations (2.8)–(2.17) of the dBWM algebra. Almost all relations immediately follow. To prove the few non-trivial cases, we use the following identities:

$$\sum_\nu G_{a,\nu}\frac{[a_\mu+a_\nu-2\lambda+1]}{[a_\mu+a_\nu+1]} = \frac{[2a_\mu-2\lambda+1]}{[2a_\mu+1]} \qquad \mu\neq 0 \tag{3.12}$$

$$\sum_\mu G_{a,\mu} = \frac{[1-2\sigma\lambda][\frac{1}{2}+\sigma\lambda]}{[\frac{1}{2}-\sigma\lambda][1]} = \sqrt{Q}. \tag{3.13}$$

Since the proof of these identities is rather cumbersome, we defer it till the appendix.

# 4 Baxterization of the dBWM algebra

We now proceed to employ the representations of the dBWM algebra as listed in the previous section to construct solutions of the YBE.

For this purpose we introduce the face operators $X_j$, $j=1,\ldots,N$ depending on a *spectral* parameter $u$, and acting in the path space $\mathcal{H}_N$. Our aim is to find face operators that form a Yang–Baxter algebra (YBA) [1]. That is, the $X_j$ satisfy the YBE

$$X_{j+1}(u)X_j(u+v)X_{j+1}(v) = X_j(v)X_{j+1}(u+v)X_j(u) \tag{4.1}$$

[3] In the sense of the vertex–RSOS duality for solvable models.

as well as the commutation relation

$$X_j(u)X_k(v) = X_k(v)X_j(u) \qquad \text{for } |j-k| \geq 2. \tag{4.2}$$

From the defining relations (2.8)–(2.17) it follows that any representation of the dBWM algebra gives rise to a YBA via [10]

$$\begin{aligned} X_j(u) \;&=\; (\,\rangle\langle\,)_j \;-\; \frac{[u]}{2\,\mathrm{i}\,[\lambda][1]}\Big(q^{u-\lambda}(\times)_j - q^{-u+\lambda}(\times)_j\Big) \\ &+\; \frac{[u][\lambda-u]}{[\lambda][1]}\Big((\,/\,)_j + (\,\backslash\,)_j\Big) \;+\; \frac{[\lambda-u]}{[\lambda]}\Big((\,\rangle\;\,)_j + (\;\,\langle\,)_j\Big) \\ &+\; \frac{[u]}{[\lambda]}\Big((\,\frown\,)_j + (\,\smile\,)_j\Big) \;+\; \frac{[2\lambda-u][\lambda+u]-\sqrt{Q}\,[\lambda-u][u]}{[2\lambda][\lambda]}\,(\quad)_j\,. \end{aligned} \tag{4.3}$$

Note that we do not require face operators at the same site to commute, this being an unnecessary condition to obtain commuting transfer matrices, see equation (4.2). In fact, our solutions do *not* have this property, i.e.,

$$X_j(u)X_j(v) \;\neq\; X_j(v)X_j(u)\,. \tag{4.4}$$

This can be seen as follows. Using expression (3.9) for $\sqrt{Q}$, equation (2.16) relating $e_j$ and $b_j^{\pm}$, and the definitions in (2.18), we get

$$\begin{aligned} X_j(\lambda) \;&=\; (\,{}^{\smile}_{\frown}\,)_j \;+\; (\,\smile\,)_j \;+\; (\,\frown\,)_j \;+\; (\quad)_j \;=\; E_j \\ q^{\mp 1}\lim_{u\to\pm\mathrm{i}\infty}\frac{X_j(u)}{\rho(u)} \;&=\; b_j^{\pm} \;+\; (\,/\,)_j \;+\; (\,\backslash\,)_j \;+\; \sigma\,(\quad)_j \;=\; B_j^{\pm}\,, \end{aligned} \tag{4.5}$$

with $\rho$ given by

$$\rho(u) \;=\; \frac{[1-u][\lambda-u]}{[1][\lambda]}. \tag{4.6}$$

Hence, in comparison with the Baxterization of the ordinary BWM algebra, the operators $B_j^{\pm}$ and $E_j$ and not $b_j^{\pm}$ and $e_j$ play the role of the braid and Temperley–Lieb operators, respectively. From (2.21) we find that $B_j$ and $E_j$ do not commute since, from eq. (3.9), we never have $\omega = \sigma$. Therefore the face operators $X_j(u)$ do not form a commuting family, establishing equation (4.4).

To prove the above Baxterization, we substitute the form (4.3) for $X_j(u)$ into the YBE (4.1). Applying the relations (2.11)–(2.14) of the dBWM algebra, and substituting relation (2.17) for $\sqrt{Q}$ as a function of $\omega$, we can, using some simple trigonometric identities, simplify to yield the cubic (2.15). The proof of the commutation relation (4.2) is trivial. Since the operators $\mathcal{O}_j$ constituting the dBWM algebra act non-trivially only at positions $j$ and $j+1$, we immediately establish the wanted result.

If we now exploit the $X_n^{(1)}$ representations (3.10) of the dBWM algebra, together with the Baxterization (4.3) we obtain three infinite families of YBA's or, equivalently, via equation (5.1) given below, three series of interactions round a face models of the RSOS type. So far, we have labelled these by the affine Lie algebras $B_n^{(1)}$, $C_n^{(1)}$ and $D_n^{(1)}$. A somewhat closer inspection reveals however that the models are nothing but RSOS counterparts [28] of the $D_{n+1}^{(2)}$, $A_{2n}^{(2)}$ and $B_n^{(1)}$ vertex models of ref. [2]. In the case of $A_{2n}^{(2)}$, the RSOS models have been obtained previously in [8] by very similar methods. The $B_n^{(1)}$ and $D_{n+1}^{(2)}$ models are new however, as it is noted that the above presented $B_n^{(1)}$ model does not coincide with the $B_n^{(1)}$ model of ref. [4], which, at criticality, is based on the ordinary BWM algebra, and not on its dilution. This occurrence of two RSOS counterparts to one and the same vertex model is similar to that as found for the $A_n^{(2)}$ models [8, 6, 7, 5].

# 5 Elliptic $A_{2n}^{(2)}$, $B_n^{(1)}$ and $D_{n+1}^{(2)}$ RSOS models

In the previous section we have shown how a given representation of the dBWM algebra can be Baxterized to yield a solution to the YBE. We now list the Boltzmann weights, defined via

$$(X_j)_{\{a\},\{b\}} = W\begin{pmatrix} a_{j-1} & b_j \\ a_j & a_{j+1} \end{pmatrix} \prod_{k\neq j} \delta_{a_k,b_k}, \tag{5.1}$$

arising from the $X_n^{(1)}$ dBWM representations (3.10) and the Baxterization (4.3). However, instead of presenting the weights in their trigonometric form as follows from the dBWM algebra, we give generalized weights involving elliptic functions. These more general weights still satisfy the YBE, but not longer have a dBWM structure.

First we need some more definitions. The elliptic theta functions $\vartheta_1$ and $\vartheta_4$ of nome $p = \exp(\pi\tau\,\mathrm{i}\,)$, Im $\tau > 0$ are defined by the following infinite products

$$\begin{aligned} \vartheta_1(u,p) &= 2p^{1/4}\sin u \prod_{n=1}^{\infty}\left(1-2p^{2n}\cos 2u + p^{4n}\right)\left(1-p^{2n}\right) \\ \vartheta_4(u,p) &= \prod_{n=1}^{\infty}\left(1-2p^{2n-1}\cos 2u + p^{4n-2}\right)\left(1-p^{2n}\right). \end{aligned} \tag{5.2}$$

With these two functions we define

$$[u] = \vartheta_1\left(\frac{s\pi u}{L},p\right) \qquad\qquad [u]_4 = \vartheta_4\left(\frac{s\pi u}{L},p\right). \tag{5.3}$$

Note that the above form for [.] replaces the earlier definition in (3.7), to which it (apart from an irrelevant factor $2p^{1/4}$) reduces in the $p \to 0$ limit.

The non-zero face weights of the three series of dilute RSOS models can now all be expressed in one unified form as

$$W\begin{pmatrix} a & a+\epsilon_\mu \\ a+\epsilon_\mu & a+2\epsilon_\mu \end{pmatrix} = \frac{[\lambda-u][1-u]}{[\lambda][1]} \qquad \mu\neq 0$$

$$W\begin{pmatrix} a & a+\epsilon_\mu \\ a+\epsilon_\mu & a+\epsilon_\mu+\epsilon_\nu \end{pmatrix} = \frac{[a_\mu - a_\nu + u][\lambda - u]}{[a_\mu - a_\nu][\lambda]} \qquad \mu \neq \pm\nu$$

$$W\begin{pmatrix} a & a+\epsilon_\nu \\ a+\epsilon_\mu & a+\epsilon_\mu+\epsilon_\nu \end{pmatrix} = \left(\frac{[a_\mu - a_\nu + 1][a_\mu - a_\nu - 1]}{[a_\mu - a_\nu]^2}\right)^{1/2} \frac{[\lambda - u][u]}{[\lambda][1]} \qquad \mu \neq \pm\nu$$

$$W\begin{pmatrix} a & a+\epsilon_\nu \\ a+\epsilon_\mu & a \end{pmatrix} = (G_{a,\mu}G_{a,\nu})^{1/2} \frac{[a_\mu + a_\nu - \lambda + 1 + u][u]}{[a_\mu + a_\nu + 1][\lambda]} \qquad \mu \neq \nu$$

$$\begin{aligned} W\begin{pmatrix} a & a+\epsilon_\mu \\ a+\epsilon_\mu & a \end{pmatrix} &= \frac{[2a_\mu + 1 + u][\lambda - u]}{[2a_\mu + 1][\lambda]} + G_{a,\mu}\frac{[2a_\mu - \lambda + 1 + u][u]}{[2a_\mu + 1][\lambda]} \qquad \mu \neq 0 \\ &= \frac{[2a_\mu - 2\lambda + 1 + u][\lambda + u]}{[2a_\mu - 2\lambda + 1][\lambda]} - H_{a,\mu}\frac{[2a_\mu - \lambda + 1 + u][u]}{[2a_\mu - 2\lambda + 1][\lambda]} \end{aligned} \qquad (5.4)$$

$$W\begin{pmatrix} a & a+\epsilon_\mu \\ a & a+\epsilon_\mu \end{pmatrix} = W\begin{pmatrix} a & a \\ a+\epsilon_\mu & a+\epsilon_\mu \end{pmatrix} = \left(\frac{[a_\mu + \frac{3}{2}]_4[a_\mu - \frac{1}{2}]_4}{[a_\mu + \frac{1}{2}]_4{}^2}\right)^{1/2} \frac{[\lambda - u][u]}{[\lambda][1]}$$

$$W\begin{pmatrix} a & a+\epsilon_\mu \\ a & a \end{pmatrix} = W\begin{pmatrix} a & a \\ a+\epsilon_\mu & a \end{pmatrix} = (G_{a,\mu})^{1/2} \frac{[a_\mu - \lambda + \frac{1}{2} + u]_4[u]}{[a_\mu + \frac{1}{2}]_4[\lambda]}$$

$$W\begin{pmatrix} a & a \\ a & a+\epsilon_\mu \end{pmatrix} = W\begin{pmatrix} a+\epsilon_\mu & a \\ a & a \end{pmatrix} = \frac{[a_\mu + \frac{1}{2} - u]_4[\lambda - u]}{[a_\mu + \frac{1}{2}]_4[\lambda]}$$

$$W\begin{pmatrix} a & a \\ a & a \end{pmatrix} = \frac{[2\lambda - u][\lambda + u]}{[2\lambda][\lambda]} - \tilde{H}_{a,0}\frac{[\lambda - u][u]}{[2\lambda][\lambda]},$$

with $\mu, \nu = \pm 1, \ldots, \pm n$ for $\mathrm{A}_{2n}^{(2)}$ and $\mathrm{B}_n^{(1)}$, and $\mu, \nu = 0, \pm 1, \ldots, \pm n$ for $\mathrm{D}_{n+1}^{(1)}$, and $a_0 = -\frac{1}{2}$.

The function $G_{a,\mu}$ is as defined in (3.11), where we of course now interpret [.] as theta function. The function $h$ therein however generalizes to

$$h(a) = \begin{cases} [2a]/[a]_4 & \text{for } \mathrm{A}_{2n}^{(2)} \\ [a]_4 & \text{for } \mathrm{B}_n^{(1)} \\ [a][a]_4 & \text{for } \mathrm{D}_{n+1}^{(2)}. \end{cases} \qquad (5.5)$$

Also the function $H_{a,\mu}$ in (3.11) changes, and we have one extra function $\tilde{H}_{a,0}$, reducing to $\sqrt{Q}$ in the $p \to 0$ limit,

$$\begin{aligned} H_{a,\mu} &= \sum_{\nu \neq \mu} G_{a,\nu}\frac{[a_\mu + a_\nu - 2\lambda + 1]}{[a_\mu + a_\nu + 1]} + \frac{[a_\mu - 2\lambda + \frac{1}{2}]_4}{[a_\mu + \frac{1}{2}]_4} \\ \tilde{H}_{a,0} &= \sum_{\mu} G_{a,\mu}\frac{[a_\mu - 2\lambda + \frac{1}{2}]_4}{[a_\mu + \frac{1}{2}]_4}. \end{aligned} \qquad (5.6)$$

The proof that the above weights satisfy the YBE will be omitted as it is completely analogous, though considerably more lengthy, to that given in [4] for the $\mathrm{X}_n^{(1)}$ RSOS models.

Instead, in the appendix we just point out the equivalence of the two different forms for the weight $W\begin{pmatrix} a & a+\epsilon_\mu \\ a+\epsilon_\mu & a \end{pmatrix}$ in the $\mu \neq 0$ case, as it also provides the proof for the identity (3.12) used in proving the dBWM representations (3.10).

As is maybe not immediately clear from equation (5.4), the $D^{(2)}_{n+1}$ Boltzmann weights exhibit an interesting symmetry. To be more precise, if we consider the unrestricted $D^{(2)}_{n+1}$ SOS model, (for the precise definition of RSOS versus SOS face models, see e.g., [3, 4]) and perform the imaginary transformation $a_\mu \to a_\mu + \frac{1}{2}\pi\tau$, for all $\mu = 1, 2, \ldots, n$, the model remains invariant by interchanging the (unordered) pairs $(a, a) \leftrightarrow (a, a+\epsilon_0)$. For example, the weight $W\begin{pmatrix} a & a \\ a & a \end{pmatrix}$ transforms to $W\begin{pmatrix} a & a+\epsilon_0 \\ a+\epsilon_0 & a \end{pmatrix}$ and vice versa, as can easily be seen by noting that under the imaginary transformation the functions $\tilde{H}_{a,0}$ and $H_{a,0}$ transform into each other (apart from an irrelevant gauge factor).

Besides the YBE for IRF models [1],

$$\sum_g W\left(\begin{matrix} f & g \\ a & b \end{matrix}\middle| u\right) W\left(\begin{matrix} e & d \\ f & g \end{matrix}\middle| u+v\right) W\left(\begin{matrix} d & c \\ g & b \end{matrix}\middle| v\right)$$

$$= \sum_g W\left(\begin{matrix} e & g \\ f & a \end{matrix}\middle| v\right) W\left(\begin{matrix} g & c \\ a & b \end{matrix}\middle| u+v\right) W\left(\begin{matrix} e & d \\ g & c \end{matrix}\middle| u\right), \tag{5.7}$$

the face weights (5.4) satisfy several standard properties, some of which are listed below,

$$W\left(\begin{matrix} d & c \\ a & b \end{matrix}\middle| 0\right) = \delta_{a,c} \qquad \text{initial condition}$$

$$W\left(\begin{matrix} d & c \\ a & b \end{matrix}\middle| \lambda - u\right) = \left(\frac{S_a S_c}{S_b S_d}\right)^{1/2} W\left(\begin{matrix} c & b \\ d & a \end{matrix}\middle| u\right) \qquad \text{crossing symmetry} \tag{5.8}$$

$$\sum_g W\left(\begin{matrix} d & g \\ a & b \end{matrix}\middle| u\right) W\left(\begin{matrix} d & c \\ g & b \end{matrix}\middle| -u\right) = \rho(u)\rho(-u)\delta_{a,c} \qquad \text{inversion relation.}$$

The *crossing multipliers* $S_a$ are given by

$$S_a = \prod_{i=1}^{n} \sigma^{a_i} h(a_i) \prod_{1\le i<j\le n} [a_i + a_j][a_i - a_j] \tag{5.9}$$

and the function $\rho$ by (4.6), interpreting [.] therein as an elliptic function. It is interesting to note that, as a consequence of (3.13), the following eigenvalue equation holds in the critical $p = 0$ limit:

$$\sum_{a\sim b} S_a = \sqrt{Q}\, S_b. \tag{5.10}$$

As shown in ref. [18], such an equation can be taken as the starting point in finding graphs which are different from the ones defined in section 3, on which $X^{(1)}_n$ dBWM representations can be built.

# 6 Summary and discussion

In this paper we have constructed several series of solvable RSOS models based on the dilute BWM algebra. These series, labelled by the affine Kac–Moody algebras $\mathrm{D}_{n+1}^{(2)}$, $\mathrm{A}_{2n}^{(2)}$ and $\mathrm{B}_{n}^{(1)}$, can naturally be viewed as dilute generalizations of the $\mathrm{B}_{n}^{(1)}$, $\mathrm{C}_{n}^{(1)}$ and $D_{n}^{(1)}$ RSOS models of Jimbo, Miwa and Okado [4], which are all based on the ordinary non-dilute BWM algebra.

Of the three dilute series, those labelled by $\mathrm{D}_{n+1}^{(2)}$ and $\mathrm{B}_{n}^{(1)}$ are new, and the latter provides an example of the possible existence of several RSOS counterparts to one and the same vertex model. The dilute $\mathrm{A}_{2n}^{(2)}$ models are the same as those found in ref. [8] and, similar to the two series of $\mathrm{B}_{n}^{(1)}$ models, are different to the $\mathrm{A}_{2n}^{(2)}$ models of ref. [5].

Though the dBWM algebra structure of the models only holds for Boltzmann weights parametrized in terms of trigonometric functions, we also presented elliptic face weights which still obey the Yang–Baxter relation.

Before we conclude, we would like to point out several possible generalizations to the working described in this paper.

Firstly, both the ordinary BWM algebra as well as the dBWM algebra can be viewed as special cases of a more general multi–colour braid–monoid algebra [21]. A natural question therefore is whether we cannot Baxterize this multi–colour algebra and find, similar to the BWM and dBWM case, level-$l$ $\mathrm{X}_{n}^{(1)}$ representations. So far we have not succeeded to do so in general, but for the case of two colours we indeed found a Baxterization which, in some sense, incorporates both the BWM and dBWM Baxterizations.

Secondly, the simplest and best well-known series of RSOS models, the ABF models [3], can be viewed as either a BWM model or a Temperley–Lieb model. Viewing it as a BWM model, it can readily be generalized to general rank to yield the $\mathrm{C}_{n}^{(1)}$ models of ref. [4]. Viewing it as a TL model, it admits an altogether different generalization to the $\mathrm{A}_{n}^{(1)}$ models of [4], which for $n > 1$ have a Hecke algebra structure. One typical feature of the latter models is that they do not possess the usual crossing symmetry. For small values of the rank $n$, we have now been able to find a non-crossing symmetric braid–monoid type algebra which gives $\mathrm{A}_{n}^{(1)}$ type RSOS models different to those of ref. [4]. Whether we can generalize this to arbitrary rank remains as yet an open question.

Finally, so far we have found dilute generalizations to all RSOS models based on the classical non-twisted affine Lie algebras. Whether also some of the exceptional algebra models admit dilution remains an open question.

We hope to report on some of the abovementioned generalizations in future publications.

# Acknowledgements

We wish to thank Paul Pearce for stimulating discussions and David O'Brien for helpful suggestions to improve the manuscript. This work has been supported by Stichting voor

Fundamenteel Onderzoek der Materie (FOM) and the Australian Research Council (ARC).

# A Proof of some identities

In this appendix we prove the two identities (3.12) and (3.13), as well as the equivalence of the two different forms for the face weight $W\begin{pmatrix} a & a+\epsilon_\mu \\ a+\epsilon_\mu & a \end{pmatrix}$ in the $\mu \neq 0$ case. In fact, proving the latter immediately leads to (3.12). We remind the reader that we are only concerned about the respective identities in the unrestricted case, where they hold irrespective of the value of $L$. From the actual form of the function $G_{a,\mu}$ and the definition of $L$ in (3.2), it is however trivial to also assert the finite forms.

## A.1 Proof of (3.12)

Before we start proving the relation (3.12) and the equivalence of the two forms in (5.4) for $W\begin{pmatrix} a & a+\epsilon_\mu \\ a+\epsilon_\mu & a \end{pmatrix}$, let us recall that in (3.12) [.] stands for the ordinary trigonometric sine function, whereas in (5.4) [.] has to be interpreted as a $\vartheta_1$-function, see (3.7) and (5.2), respectively.

As a first step in our proof, we take the difference of the two equivalent expressions in (5.4) and substitute (5.6) for $H_{a,\mu}$. Now using the standard theta function identities

$$\begin{aligned} &[u+x][u-x][v+y][v-y]-[u+y][u-y][v+x][v-x] \\ &\qquad\qquad = [x+y][x-y][u+v][u-v] \\ &[u+x][u-x][v+y]_4[v-y]_4-[u+y][u-y][v+x]_4[v-x]_4 \qquad\qquad (A.1) \\ &\qquad\qquad = -[x+y][x-y][u+v]_4[u-v]_4 \end{aligned}$$

to simplify, we are led to prove

$$\sum_\nu G_{a,\nu}\frac{[a_\mu+a_\nu-2\lambda+1]}{[a_\mu+a_\nu+1]} = \frac{[2a_\mu-2\lambda+1][a_\mu+\lambda+\frac{1}{2}]_4}{[2a_\mu+1][a_\mu-\lambda+\frac{1}{2}]_4} \qquad \mu\neq 0. \tag{A.2}$$

Before we proceed, we remark that in letting the nome $p\to 0$, which has the effect of replacing $\vartheta_4$ by 1 and $\vartheta_1$ by the sine function, the above expression reduces to the identity (3.12).

Following the method of proof of [4], we consider the function

$$f(z) = \frac{[z+a_\mu-2\lambda+1][2z]}{[z+a_\mu+1][2z+1]}\,\frac{h(z+1)}{h(z)}\prod_{\nu\neq 0}\frac{[z-a_\nu+1]}{[z-a_\nu]} \qquad \mu\neq 0. \tag{A.3}$$

From the quasi-periodicity properties

$$\begin{aligned} &[u+L/s] = -[u] \qquad && [u+\tau L/s] = -p^{-1}\mathrm{e}^{-2\mathrm{i}s\pi u/L}[u] \\ &[u+L/s]_4 = [u]_4 \qquad && [u+\tau L/s]_4 = -p^{-1}\mathrm{e}^{-2\mathrm{i}s\pi u/L}[u]_4 \end{aligned} \tag{A.4}$$

we see that $f$ is doubly periodic. Hence the contour integral along a period-parallelogram vanishes, and we have $\sum \mathrm{Res}\, f(z) = 0$. Locating the poles at $z = a_\nu$ ($\nu \neq 0$), $-\frac{1}{2}$, $-\frac{1}{2}(1 + L/s)$, $-\frac{1}{2}(1+\tau L/s)$, $-\frac{1}{2}(1+L/s+\tau L/s)$, and computing the sum over the residues, yields[4]

$$2\sum_\nu \frac{[a_\mu + a_\nu - 2\lambda + 1]}{[a_\mu + a_\nu + 1]} G_{a,\nu}$$
$$= \frac{[a_\mu - 2\lambda + \frac{1}{2}]}{[a_\mu + \frac{1}{2}]} + \frac{[a_\mu - 2\lambda + \frac{1}{2}]_2}{[a_\mu + \frac{1}{2}]_2} + \frac{[a_\mu - 2\lambda + \frac{1}{2}]_3}{[a_\mu + \frac{1}{2}]_3} - \frac{[a_\mu - 2\lambda + \frac{1}{2}]_4}{[a_\mu + \frac{1}{2}]_4}. \quad \text{(A.5)}$$

Here $[u]_2 = [u + L/(2s)]$, $[u]_3 = [u + L/(2s)]_4$, $[u + \tau L/(2s)] = \mathrm{i}\, p^{-1/4} \exp(-\mathrm{i}\, s\pi u/L)[u]_4$ and $\mu \neq 0$. Repeated use of the addition formulae (A.1) yields

$$\frac{[u+x]}{[u-x]} + \frac{[u+x]_2}{[u-x]_2} + \frac{[u+x]_3}{[u-x]_3} - \frac{[u+x]_4}{[u-x]_4} = 2\, \frac{[2u][u-2x]_4}{[2u-2x][u]_4}. \quad \text{(A.6)}$$

Setting $u = a_\mu - \lambda + \frac{1}{2}$ and $x = -\lambda$ in this relation, we find that (A.5) reduces to (A.2), which completes the proof.

## A.2 Proof of (3.13)

To establish identity (3.13), we adopt a somewhat similar method of proof as in proving (3.12). That is, although (3.13) holds for [.] being a trigonometric function only, we start by considering the following expression, where [.] and $[.]_4$ are elliptic functions as defined in (5.2):

$$g(z) = \frac{[z + a_\mu - 2\lambda + 1]_4 [2z]}{[z + a_\mu + 1]_4 [2z+1]}\, \frac{h(z+1)}{h(z)} \prod_{\nu \neq 0} \frac{[z - a_\nu + 1]}{[z - a_\nu]} \qquad \mu \neq 0. \quad \text{(A.7)}$$

We again find double periodicity and hence that the sum over the residues within a period-parallelogram must vanish. The poles of $g$ are exactly those of the function $f$ in (A.3), plus one additional pole at $z = -a_\mu - 1 + \tau L/(2s)$. Performing the sum $\sum \mathrm{Res}\, g(z)$ yields

$$\sum_\nu \frac{[a_\mu + a_\nu - 2\lambda + 1]_4}{[a_\mu + a_\nu + 1]_4} G_{a,\nu}$$
$$= \sigma \frac{\tilde{h}(a_\mu)}{\tilde{h}(a_\mu + 1)} \frac{[2\lambda][2a_\mu + 2]}{[1][2a_\mu + 1]} \prod_{\nu \neq 0} \frac{[a_\mu - a_\nu]_4}{[a_\mu - a_\nu + 1]_4}$$
$$-\frac{1}{2}\left( \frac{[a_\mu - 2\lambda + \frac{1}{2}]}{[a_\mu + \frac{1}{2}]} + \frac{[a_\mu - 2\lambda + \frac{1}{2}]_2}{[a_\mu + \frac{1}{2}]_2} + \frac{[a_\mu - 2\lambda + \frac{1}{2}]_3}{[a_\mu + \frac{1}{2}]_3} + \frac{[a_\mu - 2\lambda + \frac{1}{2}]_4}{[a_\mu + \frac{1}{2}]_4} \right) \quad \text{(A.8)}$$
$$= \sigma \frac{\tilde{h}(a_\mu)}{\tilde{h}(a_\mu + 1)} \frac{[2\lambda][2a_\mu + 2]}{[1][2a_\mu + 1]} \prod_{\nu \neq 0} \frac{[a_\mu - a_\nu]_4}{[a_\mu - a_\nu + 1]_4} + \frac{[2a_\mu - 2\lambda + 1][a_\mu + \lambda + \frac{1}{2}]}{[2a_\mu + 1][a_\mu - \lambda + \frac{1}{2}]},$$

[4]For $\mathrm{D}^{(2)}_{n+1}$, the term $\nu = 0$ in the lhs of (A.5) arises from the fact that the pole at $z = -\frac{1}{2}$ in fact yields $-[a_\mu - 2\lambda + \frac{1}{2}]/[a_\mu + \frac{1}{2}]$ in the rhs. Adding the term $\nu = 0$ in the sum, recalling that $G_{a,0} = 1$, leads to (A.5).

where in the last step we have used (A.6) with $u = a_\mu - \lambda + \frac{1}{2} + \tau L/(2s)$ and $x = -\lambda$. The function $\tilde{h}$ is given by

$$\tilde{h}(a) = \begin{cases} [2a]/[a] & \text{for } \mathrm{A}_{2n}^{(2)} \\ [a] & \text{for } \mathrm{B}_{n}^{(1)} \\ [a][a]_4 & \text{for } \mathrm{D}_{n+1}^{(2)}. \end{cases} \tag{A.9}$$

We now take the nome $p$ to zero limit, replacing $[.]_4$ by 1 and reinterpreting $[.]$ as the trigonometric function (3.7). This removes the complicated product in the right hand side and allows simplification to

$$\sum_\nu G_{a,\nu} = \frac{[1-2\sigma\lambda][\frac{1}{2}+\sigma\lambda]}{[\frac{1}{2}-\sigma\lambda][1]}. \tag{A.10}$$

This is precisely the identity we were after and hence we are done.